\documentclass[a4paper,10pt]{article}
\usepackage{graphicx}

\title{RXTE-ASM light curves of Cen X-3}
\author{Biswajit Paul and Harsha Raichur\\
Raman Research Institute, Bangalore 560080, India}

\date{}

\begin{document}

\maketitle

\begin{abstract}
In Paul, Raichur and Mukherjee (2005) we had reported a spectral mode change in the X-ray binary pulsar Cen X-3
using the then available ASM light curves of Cen X-3. Being mindful of the fact that instrument calibration or
other issues like gain changes, software issues etc. might give rise to some of the artifacts, we had used the ASM light 
curves of three other X-ray binary pulsars namely Her X-1, SMC X-1 and Vela X-1 for comparison for the same observation times. 
In light of the recent finding of M\"{u}ller et al. (2011); which reports the non-detection of such features in the current 
ASM light curves available on the HEASARC website and with other instruments like MAXI, SWIFT and INTEGRAL, we would like 
to point out the inconsistency between the ASM ligthcurves available in 2005 and now. This information will be very useful 
for other users of ASM light curves.
\end{abstract}

\section*{Introduction}

Using multi-band light curve of the RXTE-ASM, we had reported a 
significant change in the X-ray spectrum of Cen X-3 in the period 
December 2000 to April 2004. This was interpreted as a change in 
the accretion mode in this period (Paul, Raichur \& Mukherjee, 2005). 
Recently, M\"{u}ller et al. (2011) reported non detection of this feature 
in the RXTE-ASM light curve of the same time span and also in some 
additional data from RXTE-ASM, MAXI, Integral and Swift. We fully 
agree with the findings of M\"{u}ller et al. (2011)

Here we report that the earlier finding was not due to any error in 
analysis of the light curves used in Paul, Raichur \& Mukherjee (2005) 
but due to a discrepancy in the ASM light curve for the same period 
that was made available in 2005 and what is available now.

\section*{RXTE-ASM light curves}
We would like to note here that the files used by us for the report of 
2005 were all for observations from 4th or 5th January 1996 to 
7th November 2004 and all the fits files were created on 9th 
November 2004. 

Below we show a plot of the light curves in the three energy bands that was available in 2005
(with no further analysis done on the original lightcurves) and the light curves now available from 
the NASA's High energy Astrophysics Science Archive Center (HEASARC). 
All these light curves are the dwell-by-dwell data as used in the M\"{u}ller et al (2011) research note. 
The time bin used to plot the light curves is 2.08702 d. The red points
are from the older light curves and the black points are from the current light curves.
Only data points upto TJD 13600 are ploted for clarity. While small differences are seen between the
two sets for the 1.5-3 keV and 3-5 keV bands, the two light curves in the 5-12 keV band are
significantly different in the second half of the plot.
We have also verified that this difference is present only for the light curves of Cen X-3 and 
not in a significant way for the other sources namely Vela X-1, Her X-1 and SMC X-1 
for which we have the older version of light curves to compare. 
The older version of the light curves for Cen X-3, Her X-1, Vela X-1
and SMC X-1 are available with the authors and can be shared on request.

\begin{figure*}[h]
\includegraphics[height=90mm,angle=-90]{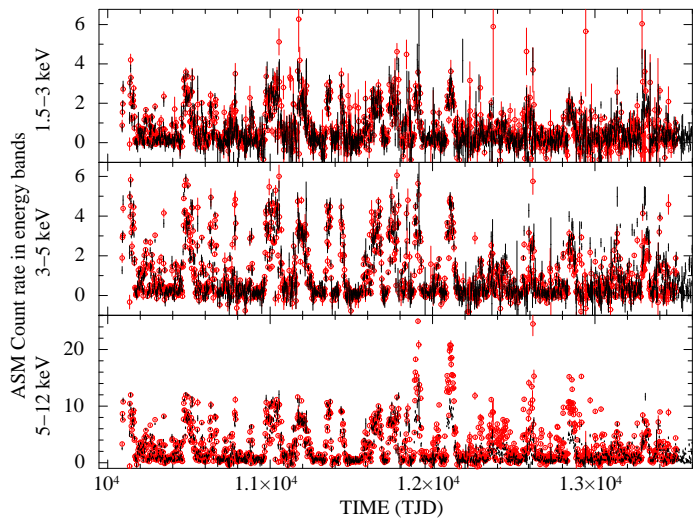}
\caption{The light curves of Cen X-3 are shown here with a binsize of 2.087 days.}
\end{figure*}

\section*{Conclusion}
The multiband RXTE-ASM light curves of Cen X-3 that were available at HEASARC in 2005
are different from the light curves that are now available at the same site.
We would also like to note here that before publishing the results in 2005, we had discussed
with the RXTE-ASM team members about any possible artifacts. The difference in the light
curves was noticed in 2010 and the same was communicated to the RXTE-ASM team. 
As yet, no clarification is available. This information about possible artifacts will be
useful for users of ASM light curves.

\end{document}